# All-optical band structure reconstruction and onset of Landau quantization of Dirac fermions

J. Riepl[1,+], M. Aichner[1,2,+], N. N. Mikhailov[3], S. A. Dvoretsky[3], G. Budkin[4], S. Ganichev[1], C. Lange[5], J. Mornhinweg[1,5,6,*] and R. Huber[1]

[1]Dept. of Physics, University of Regensburg, 93040 Regensburg, Germany

[2]LSI, CNRS, CEA/DRF/IRAMIS, Ecole Polytechnique, Institut Polytechnique de Paris, F-91120 Palaiseau, France

[3]A.V. Rzhanov Institute of Semiconductor Physics, Novosibirsk 630090, Russia

[4]A. F. Ioffe Insitute, Saint Petersburg, Russia

[5]Dept. of Physics, TU Dortmund University, 44227 Dortmund, Germany

[6]Harvard John A. Paulson School of Engineering and Applied Sciences, 02138 Cambridge, MA, USA

*Corresponding author, jmornhinweg@seas.harvard.edu

[+]These authors contributed equally

**The nature of relativistic electrons in solids depends on the precise shape of the underlying band structure. Prominently, symmetry-related mechanisms, such as the breaking of time reversal symmetry in topological insulators, can lead to the emergence of band gaps on small energy scales. It is, thus, important to quantify potential gaps of the Dirac cone with meV precision. Yet established band structure measurements are often challenged by their strict surface sensitivity or limited energy resolution. In this work, we use broadband, time-resolved THz magneto-spectroscopy to access the band structure of Dirac electrons in a buried HgTe quantum well by contact-free, all-optical measurements. Optical doping allows us to control the Fermi level without applying any electrical gate voltages. The background-free measurement of the cyclotron resonance of the Dirac system over 2.5 optical octaves, a broad range of magnetic field strengths, and different Fermi energies allows us to reconstruct the band structure near the Dirac point with sub-meV precision, and to observe a crossover of Landau quantization from a quasi-classical to the relativistic regime.**

Relativistic electrons in solids enable unique electronic and optical properties, ranging from high conductivity to universal absorption coefficients and intriguing phase transitions[1,2,3]. These phenomena critically depend on the specific electronic band structure as potential energy gaps of the Dirac cone could strongly modify the nature of the quasi-relativistic states. For example, in graphene, spin-orbit interaction that breaks time-reversal symmetry can lead to an opening of minute energy gaps[4,5,6]. Unlike graphene, which has two spin-degenerate Dirac cones at two inequivalent corners of the Brillouin zone, HgTe quantum wells (QWs) of critical thickness can possess a single spin-degenerate Dirac valley around the Γ-point[2]. Advantageously, HgTe/Hg$_x$Cd$_{1-x}$Te QWs can be grown on chip-sized areas with high quality[5,7,8]. Furthermore, these QWs feature strong intrinsic spin-orbit coupling and topologically protected edge states[5,7,9]. They offer the exceptional ability to custom-tailor the quasi-relativistic band structure near the Γ-point as well as the band gap by slightly varying the mercury concentration, x, or the thickness of the QW. This allows, e.g., for studying the transition from gapless to gapped Dirac states in the same sample configuration[2,10]. Yet, owing to the strong dependence of the band structure near the band gap on changes of the QW thickness on the scale of only a few Å (Ref. 5), an experimental characterization of the electronic properties remains indispensable. Since the required protective top layers of the QWs make established band structure mapping including reconstruction by angle-resolved photoemission spectroscopy (ARPES)[11], quantum twisting microscopy[12] or scanning-tunneling microscopy (STM)[13] infeasible, alternative methods need to be established. Here, we present an all-optical, contact-free way of reconstructing the band structure of Dirac materials by spectrally resolving the cyclotron resonance (CR) for different Fermi energies $E_F$.

Transport measurements are the generally established way to measure and confirm the Dirac-like band structure of HgTe QWs of critical thickness, however perturbations from electrical contacts are unavoidable[5,10]. Past studies have performed magneto-spectroscopy with HgTe QWs far away from the critical thickness[14], or investigated the Landau level splitting at the critical thickness by monochromatic THz radiation[15,16], at a fixed Fermi level[17]. Recently, the monochromatic excitation of CRs of an electrically contacted HgTe QW has been used for band structure reconstruction[18], leading to different results in the case of increasing as compared to decreasing gate voltages. Yet, reconstructing the precise band structure at the critical thickness, unperturbed from gate contacts and with sub-meV precision remains challenging.

We employ broadband THz time-domain transmission magneto-spectroscopy of multiple Landau level transitions to determine the precise band structure of a HgTe QW of critical thickness near the Γ-point. Contact-free optical doping allows us to vary the Fermi level without applying an electrostatic gate voltage. Our measurements reveal a crossover of non-relativistic to relativistic scaling of the CR frequency with decreasing Fermi energy – a clear signature of massless charge carriers. In conjunction with a model of Landau quantization of charge carriers in a mixed regime of quadratic and linear dispersion, we map out the relativistic band structure including its characteristic discontinuities, with sub-meV precision.

Our structure was grown on a (013)-oriented GaAs substrate using molecular beam epitaxy (details on growth process see Ref. 19). First, a strain-compensating buffer layer of ZnTe and CdTe was deposited, followed by the QW system consisting of two $Hg_{0.65}Cd_{0.35}Te$ barrier layers (thickness, 39 nm) with the HgTe QW of the critical thickness of 6.5 nm sandwiched in between. Finally, a protective layer with a thickness of 40 nm was added (see schematic in Fig. 1, supplementary material for details). Contact-free optical doping[14,16] allows us to vary the Fermi energy $E_F$. Alike for the persistent photoconductivity caused by ionization of the autolocalized DX and EL2 centers in II-V structures[20,21,22], visible light excites electrons from autolocalized impurities, located at $k \neq 0$ with varying energies into the conduction band, where they relax towards the band minimum at $k = 0$. At cryogenic temperatures of 4 K, momentum conservation inhibits radiative recombination of these photoinduced electrons. By heating the QW up to room temperature, the electrons gain enough thermal energy to overcome the momentum mismatch between the conduction band minimum and impurity states, allowing them to return to their bound states[16]. Even though the type of impurities resulting in a contribution to the optical doping by ionization has not yet been explored, this persistent photoconductivity effect is well established for HgTe semiconductors[16,23] (see supplementary material for details) and avoids undesired effects like, e.g., local charge inhomogeneities or hysteresis resulting from electrostatic gating solutions. To keep $E_F$ constant during the measurements, the sample is placed in a cryostat and cooled down to a lattice temperature of ~4 K. Undesired photodoping is prevented by shielding the cryostat windows from visible radiation using a black polypropylene film with a thickness of 0.4 mm and high transparency in the THz spectral range. A static magnetic bias field $B$, applied perpendicularly to the QW plane, breaks time-reversal symmetry and Landau-quantizes the system.

We measure the resulting CR of the QW by time-resolved THz magneto-spectroscopy in transmission (Fig. 1). To this end, we excite our sample with a short THz pulse ($\mathcal{E}_{x,\text{THz}}$), which is linearly polarized in $x$-direction and covers a broad spectral range of 2.5 optical octaves between 0.35 and 2.5 THz (Fig. 1, inset (i)). The circular polarization of the CR re-radiates into the far-field, where we select the $y$-polarized component using a linear polarizer (LP) and measure the resulting waveform $\mathcal{E}_{y,B}(t)$ with electro-optic detection. Birefringence of the diamond windows of the cryostat converts some of the polarization of the incident transient to the $y$-axis, which we numerically subtract by a reference measurement of $\mathcal{E}_{y,B=0\,\text{T}}$ for a magnetic field of $B = 0$ T. This way, we isolate the pure far-field signal of the CR, $\mathcal{E}_{y,B>0\,\text{T}}$ (Fig. 1, inset (iii)). Subsequently, we perform a Fourier transform of $\mathcal{E}_{y,B}(t)$ and normalize the resulting amplitude spectra to the spectrum of the incident THz pulse. The resulting data allow the extraction of the CR frequencies as well as their amplitudes as a function of the magnetic field $B$. Optical doping is achieved by irradiating the QW with white light (exposure time $t_e$; spectrum, see supplementary material) from a light-emitting diode positioned between the plastic film and the cryostat window.

Figure 2, left column, shows the resulting transmission spectra as a function of the magnetic field $B$ and frequency $\nu$, for different exposure times $t_e$. For $t_e = 4.01$ s and magnetic fields $B > 0.1$ T, the spectra exhibit a single maximum at a frequency $\nu_c(B)$ which increases linearly with increasing $B$ (Fig. 2(a)) and can be attributed to the CR, as shown further below. By gradually reducing $t_e$ (Fig. 2(c,e,g)), the overall slope of $\nu_c(B)$ increases. Additionally, abrupt discontinuities start to appear in the CR: for certain magnetic fields, the frequency of the CR does not rise continuously with $B$ but transitions abruptly to higher frequencies. These marked discontinuities become clearly visible in the cases of low exposure times, e.g., for $t_e = 32$ ms at $B = 0.75$ T (Fig. 2(g)).

To explain these observations, we consider the Landau splitting of systems with mixed character exhibiting both massive and Dirac electron properties. In the usual case of a parabolic dispersion relation, the Landau level energies $E_n^p = \hbar n \frac{eB}{m_e}$ increase linearly with rising magnetic field strength $B$, and fan out with equidistant spacing by their level index $n$. Here, $\hbar$, $e$, and $m_e$ are the reduced Planck constant, the elementary charge, and the effective electron mass, respectively. In contrast, the Dirac equation describes electrons following a perfectly linear dispersion relation. The resulting Landau levels exhibit a square root dependence on $B$ and $n$ according to $E_n^\lambda = \lambda v_F \sqrt{2\hbar eB|n|}$, with $\lambda = \pm 1$ denoting the conduction or valence band, respectively[24]. The Fermi velocity $v_F$ determines the slope of a purely linear band structure. Importantly, Dirac systems exhibit an extra Landau level at $E = 0$ (Ref. 25,26).

Unlike systems with parabolic band structures and small linear corrections[27,28,29], our HgTe QW features a predominantly linear band structure with minor quadratic corrections. The conduction ($E^+(k)$) and valence band ($E^-(k)$) of HgTe QWs are well described by the relation[2,3,5,10,17]

$$E^\pm(k) = \mathcal{D}k^2 \pm \sqrt{\mathcal{M}^2 - 2\mathcal{M}\mathcal{B}k^2 + \mathcal{A}^2 k^2 + \mathcal{B}^2 k^4}. \tag{1}$$

The dominant linear part is given by the material parameter $\mathcal{A}$, while quadratic corrections are accounted for by the parameters $\mathcal{D}$ and $\mathcal{B}$. The band gap $E_g$ of the QW is given by $E^+(0) - E^-(0) = 2|\mathcal{M}|$. Due to the quadratic terms and the Zeeman effect, the spin degeneracy is lifted under Landau quantization. The eigenenergies of the Landau levels are then given by[3,10]

$$E_n^{\pm,\uparrow} = -\frac{eB}{\hbar}(2\mathcal{D}n + \mathcal{B}) + \frac{g_e + g_h}{4}\mu_B B \pm \sqrt{2n\mathcal{A}^2 \frac{eB}{\hbar} + \left(\mathcal{M} - \frac{eB}{\hbar}(\mathcal{D} + 2\mathcal{B}n) + \frac{g_e - g_h}{4}\mu_B B\right)^2}, \tag{2}$$

$$E_n^{\pm,\downarrow} = -\frac{eB}{\hbar}(2\mathcal{D}n - \mathcal{B}) - \frac{g_e + g_h}{4}\mu_B B \pm \sqrt{2n\mathcal{A}^2 \frac{eB}{\hbar} + \left(\mathcal{M} - \frac{eB}{\hbar}(-\mathcal{D} + 2\mathcal{B}n) - \frac{g_e - g_h}{4}\mu_B B\right)^2}, \tag{3}$$

for $n \geq 1$ and $E_0^\uparrow = \mathcal{M} - \frac{eB}{\hbar}(\mathcal{D} + \mathcal{B}) + \frac{g_e}{2}\mu_B B$; $E_0^\downarrow = -\mathcal{M} + \frac{eB}{\hbar}(-\mathcal{D} + \mathcal{B}) - \frac{g_h}{2}\mu_B B$ for $n = 1$.

Figure 3(a) shows the eigenenergies of the first 50 Landau levels ($|n| \leq 50$) as a function of the magnetic field in the case of spin up (red) and spin down (gray) (material parameters, $\mathcal{A} = 340$ meV nm, $\mathcal{B} = -350$ meV nm$^2$, $\mathcal{D} = -780$ meV nm$^2$, $\mathcal{M} = -0.67$ meV, $g_e = 60$, $g_h = 0$). Owing to dipole selection rules, only intraband transitions between neighboring Landau levels ($\Delta n = \pm 1$) within one spin sub-system are allowed[24]. Since the sample temperature of $T \approx 4$ K corresponds to a thermal energy of 0.3 meV, it is a good approximation to assume that all Landau levels below the Fermi energy $E_\text{F}$ are fully occupied, whereas those above are empty. Therefore, only transitions between the two Landau levels $n$ and $n + 1$ enclosing the Fermi level can occur. If the magnetic field is swept within a range within which the Fermi level does not cross these levels, the CR frequency of a Dirac system is tuned according to $v_n^\lambda = \lambda v_\text{F}\sqrt{2\hbar eB}(\sqrt{|n+1|} - \sqrt{|n|})$ and a continuous spectral shape is obtained. If $E_\text{F}$ does however cross Landau levels and $n$ changes, $v_n^\lambda$ increases or decreases abruptly, leaving characteristic spectral gaps in the spectra of the magnetic field sweep. Owing to the square root dependence $v_n^\lambda \propto (\sqrt{|n+1|} - \sqrt{|n|})$, the width $\Delta v$ of these gaps decreases with increasing $n$, such that quasi-continuous spectral tuning is observed once $\Delta v$ becomes smaller than the line width of the CR. The latter situation is obtained in good approximation for $E_\text{F} = 90$ meV (Fig. 3(a), upper dashed horizontal line) and $B < 1$ T. Here, $v_n^\lambda(B)$ consists of many segments of piecewise square root form which are stitched together by small jumps $\Delta v$. In Fig. 3(a), the corresponding transitions are marked by red and gray shades for the spin-up and spin-down systems, respectively. As $B$ is increased, these jumps $\Delta v > 0$ move the sliding average of the CR frequency upwards and lead to a net linear relationship $v \propto B$ (Fig. 3(b)), which is the behavior of massive Landau-quantized electrons. On the contrary, a Fermi level of $E_\text{F} = 25$ meV (Fig. 3(a), lower dashed horizontal line) activates the lower Landau levels with small indices $n$, leading to much larger discontinuities $\Delta v$ between segments of local square root scaling, over much larger ranges of $B$ (red and gray-shaded areas). While a linear behavior of $v_n^\lambda(B)$ is observed for small magnetic fields, the discontinuous, square root-shaped pattern is clearly visible for $B \gtrsim 0.3$ T. This behavior is a hallmark of Landau level formation within a linear band structure.

Our broadband measurements do not only showcase the transition of massive, Landau-quantized electrons to their relativistic counterparts, but moreover allow us to reconstruct the band structure of the HgTe quantum well over a large energy scale. In the well-established case of a purely quadratic dispersion relation, the CR frequency grows linearly with increasing magnetic field, revealing the effective mass of the underlying charge carriers. In this case, all Landau levels are equally spaced in energy and no information about the Fermi energy of the system can be extracted directly. In contrast, the measurement of the CR of a Dirac system reveals much more details of the underlying band structure.

First, the exact locations of the discontinuities in $\nu_n^\lambda(B)$ strongly depend on the Fermi level (see Fig. 3). Additionally, the continuous spectral tuning of the CR between the frequency jumps directly marks the energy difference between two neighboring sets of Landau levels. Since this difference also depends nonlinearly on the magnetic field $B$ and the Landau level index $n$ (see Eqs. (2) and (3)), fitting the CR allows us to precisely retrieve the band structure.

To this end, we derive the dielectric function of charge carriers that generically describe the Dirac electrons in HgTe QWs (see supplementary material for details). In combination with a transmission matrix framework, we simulate the emitted polarization from the CRs for different band structure parameters ($\mathcal{A}$, $\mathcal{B}$, $\mathcal{D}$, $\mathcal{M}$) and Fermi energies $E_\text{F}$. Our fitting routine allows us to compare the calculated resonances to the measured data (see supplementary material for details). This set of material parameters has to reproduce all experimentally measured resonances (left column of Fig. 2) with the same set of parameters $\mathcal{A}$, $\mathcal{B}$, $\mathcal{D}$, $\mathcal{M}$, while only varying $E_\text{F}$. The right column of Fig. 2 ((b, d, f, h)) shows the simulated cyclotron radiation color-coded as a function of the magnetic field $B$ and frequency $\nu$ for one such set, which reproduces all experimentally measured resonances ($\mathcal{A} = 340$ meV nm, $\mathcal{B} = -350$ meV nm$^2$, $\mathcal{D} = -780$ meV nm$^2$, $\mathcal{M} = -0.67$ meV). With this fitting routine we can reconstruct the band structure of our HgTe quantum well (Fig. 4), allowing us to determine the material parameters within a small error window.

Comparing the single parameters of all accepted parameter sets leads to average values of $\mathcal{A} = 338 \pm 9$ meV nm, $\mathcal{B} = -313 \pm 72$ meV nm$^2$, $\mathcal{D} = -795 \pm 58$ meV nm$^2$, and $\mathcal{M} = -0.58 \pm 0.40$ meV. As $\mathcal{B}$ and $\mathcal{D}$ both scale the quadratic dispersion term, they are interchangeable for different sets of parameters, leading to a higher standard deviation compared to their mean value. These uncertainties for the almost linear dispersion relation result in an average standard deviation of $\sigma = 0.35$ meV in energy for the region shown in Fig. 4. Remarkably, we can even precisely map out the size of the bandgap to be $1.16 \pm 0.80$ meV (see Fig. 4(b)). To check the consistency of our results with well-established theory approaches[30], we perform a $\mathbf{k} \cdot \mathbf{p}$ – calculation of the electron spectrum for our QW using material parameters of the HgTe and CdTe listed in Ref. 31 (see supplementary material for details). Our experimentally reconstructed dispersion relation is in excellent agreement with the results of the **k · p** calculation within our standard deviation (Fig. 4, black dots).

In this paper we showed an all-optical, contact-free method to reconstruct the band structure of Dirac fermions by measuring their CR with an average accuracy of $\sigma = 0.35$ meV. By changing the Fermi level of this Dirac system, a characteristic transition of the cyclotron resonance from the quasi-classical to the relativistic state was observed. This approach allowed us to quantify small quadratic corrections to the dominantly linear dispersion relation in a HgTe quantum well of critical thickness. Our all-optical procedure adds to the methods for contact-free band structure reconstruction including materials where protective layers impede established techniques such as angle-resolved photoemission spectroscopy (ARPES) or which are highly sensitive to their dielectric environment, such as flat band structures in magic-angle graphene[32]. Non-invasively exploring these intrinsic band structure properties also opens up perspectives for on-chip optics[33,34], lightwave electronics[28,35,36,37] and quantum-information processing[38], where the knowledge of the precise band structure is vital for the control of electron dynamics.

**Supplementary Material**

Additional material on the optical doing, the sample structure, the simulations and fitting procedure as well as the $\mathbf{k} \cdot \mathbf{p}$ calculation is included in the supplementary material.

**Acknowledgments**

The authors thank Imke Gronwald for valuable discussions and technical support.

The work in Regensburg was supported by the Deutsche Forschungsgemeinschaft (DFG; German Research Foundation) through SFB 1277, project-ID 314695032 and research grant LA 3307/1-2 and HU1598/8. S.D.G. thanks the DFG for the support via project-ID 521083032 (Ga501/19). J. M. acknowledges funding from the Alexander von Humboldt foundation. G.V.B. acknowledges support by the "BASIS" foundation.

**Author declarations**

**Conflict of interest**

The authors have no conflicts to disclose.

**Data availability**

The data that supports the findings of this study are available from the corresponding author upon reasonable request.

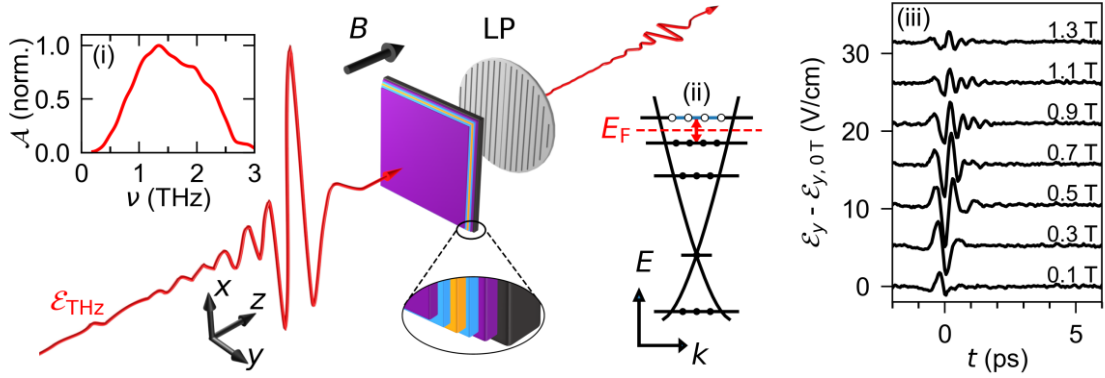

**Figure 1: Background-free detection of cyclotron resonances**. An external magnetic field $B$ is applied perpendicularly to the surface of a HgTe QW of critical thickness. The structure of the QW is shown in the zoom in. (dark gray: GaAs substrate, purple: buffer layer, light blue: HgCdTe barriers, yellow: HgTe QW, purple: capping). A THz-waveform $\mathcal{E}_{THz}$ (red) polarized in $x$-direction is focused on the QW (Inset (i): Amplitude spectrum $\mathcal{A}(\nu)$ of $\mathcal{E}_{THz}$), exciting the Landau-quantized Dirac electrons below Fermi level $E_F$ (see inset (ii)). A linear polarizer (LP) isolates the $y$-polarized field component, allowing us to measure the response of the cyclotron resonance (see inset (iii)) background-free for varying magnetic fields $B$.

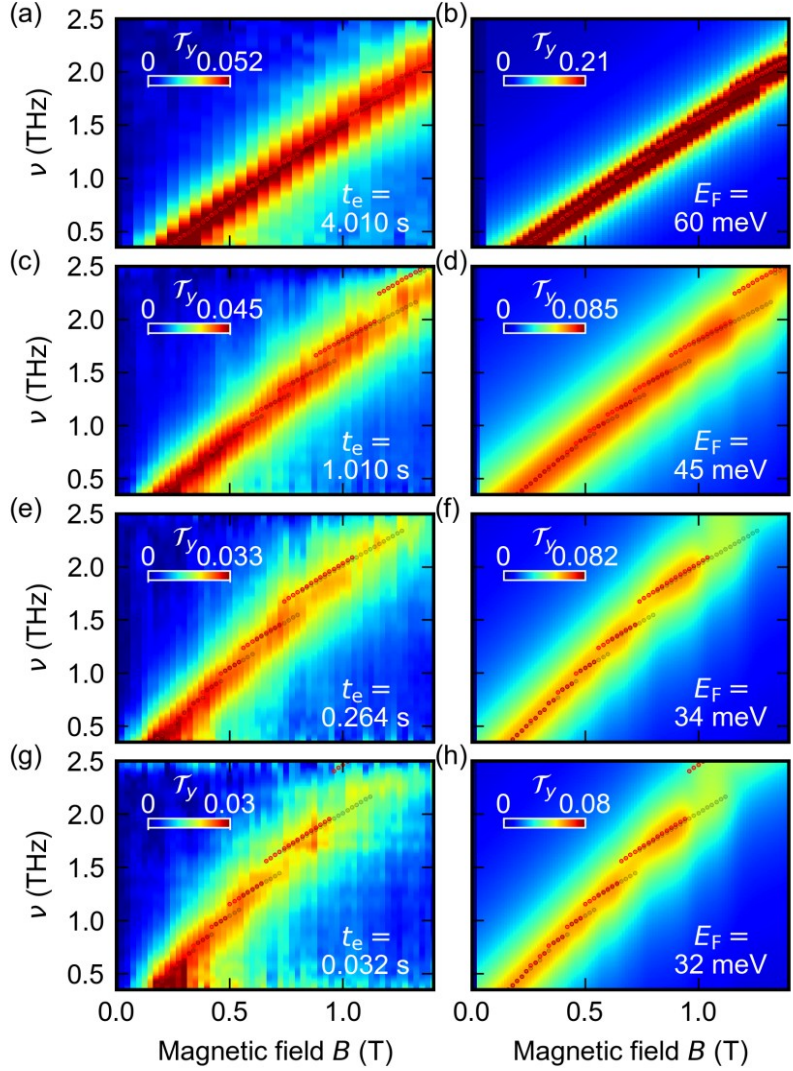

**Figure 2: Cyclotron resonance of Dirac electrons.** The measured and simulated cross polarized transmission $\mathcal{T}_y$ of the cyclotron resonance (left and right column, respectively) for different exposure times $t_e$ and fitted Fermi levels $E_F$ is shown color-coded as a function of the magnetic field $B$ and the frequency $\nu$. The red and gray dotted lines are a guide to the eye and represent the cyclotron resonance frequency for the spin up and down system, respectively. (a), (b) For a large Fermi energy of $E_F = 60.0 \pm 0.8$ meV, a quasi-classical linearly increasing cyclotron resonance is observable. (c), (d) For lower Fermi levels of $E_F = 45.5 \pm 0.5$ meV the observed slope of the resonance is rising. (e)-(h) With further decreasing Fermi levels of $E_F$ ((e,f) $E_F = 33.6 \pm 0.6$ meV, (g,h) $E_F = 32.1 \pm 0.3$ meV), jump discontinuities become visible in the cyclotron resonances of the measured and simulated spectra.

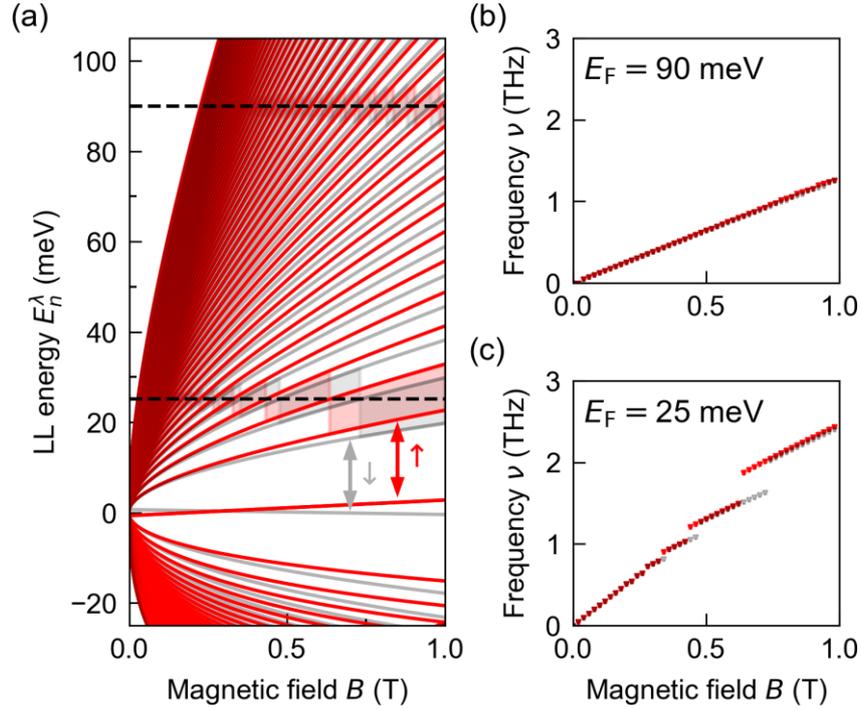

**Figure 3: Possible cyclotron transitions for different Fermi energies $E_F$.** (a) Energies of the first 50 Landau levels (red: spin up, gray: spin down) as functions of the magnetic field. For both Fermi energies, the optically active segments of the spin-up system are marked with red shading and with gray shading for the spin-down system, correspondingly. (b) In the case of a high Fermi level ($E_F$ = 90 meV), the resulting cyclotron resonance scales linearly with the magnetic field. (c) For a low Fermi energy ($E_F$ = 25 meV), discontinuities appear in the cyclotron resonance.

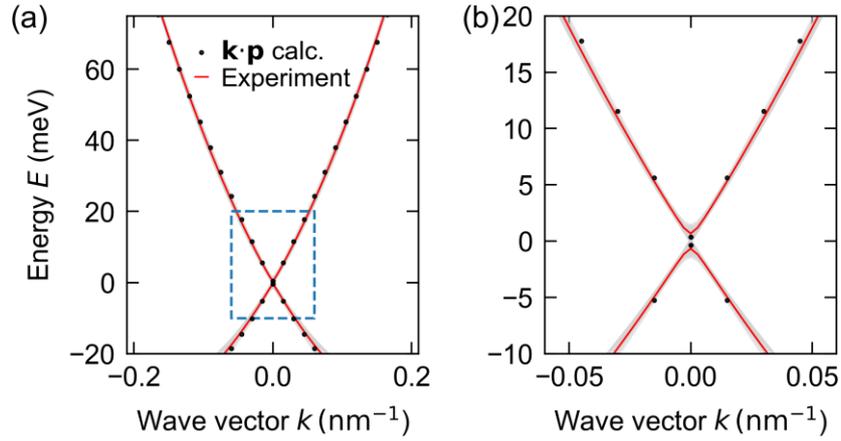

**Figure 4: Measured band structure of the HgTe QW.** (a) Band structure extracted from the best fit to the experimental cyclotron resonances shown in red. The gray region displays the error in the reconstruction procedure. Band structure predicted by $\mathbf{k} \cdot \mathbf{p}$ - calculation (black points). (b) Zoom in into the region around the Γ-point (blue rectangle in (a)). Our reconstruction predicts a bandgap smaller than 2 meV.

**References**

1. K.S. Novoselov, V.I. Fal'ko, L. Colombo, P.R. Gellert, M.G. Schwab, and K. Kim, "A roadmap for graphene," Nature **490,** 192–200 (2012).
2. B. Scharf, A. Matos-Abiague, and J. Fabian, "Magnetic properties of HgTe quantum wells," Phys. Rev. B **86** (2012).
3. S. Gebert, C. Consejo, S.S. Krishtopenko, S. Ruffenach, M. Szola, J. Torres, C. Bray, B. Jouault, M. Orlita, X. Baudry, P. Ballet, S.V. Morozov, V.I. Gavrilenko, N.N. Mikhailov, S.A. Dvoretskii, and F. Teppe, "Terahertz cyclotron emission from two-dimensional Dirac fermions," Nat. Photon. **17**, 244 (2023).
4. C.L. Kane and E.J. Mele, "Quantum spin Hall effect in graphene," Phys. Rev. Lett. **95**, 226804 (2005).
5. B.A. Bernevig, T.L. Hughes, and S.-C. Zhang, "Quantum spin Hall effect and topological phase transition in HgTe quantum wells," Science **314,** 1757 (2006).
6. J.W. McIver, B. Schulte, F.-U. Stein, T. Matsuyama, G. Jotzu, G. Meier, and A. Cavalleri, "Light-induced anomalous Hall effect in graphene," Nat. Phys. **16,** 38-4 (2020).
7. M. König, S. Wiedmann, C. Brüne, A. Roth, H. Buhmann, L.W. Molenkamp, X.-L. Qi, and S.-C. Zhang, "Quantum spin hall insulator state in HgTe quantum wells," Science **318**, 766 (2007).
8. E.B. Olshanetsky, S. Sassine, Z.D. Kvon, N.N. Mikhailov, S.A. Dvoretsky, J.C. Portal, and A.L. Aseev, "Quantum Hall liquid-insulator and plateau-to-plateau transitions in a high mobility 2D electron gas in an HgTe quantum well," JETP Lett. **84**, 565 (2007).
9. M.Z. Hasan and C.L. Kane, "Colloquium: Topological insulators," Rev. Mod. Phys. **82,** 3045 (2010).
10. B. Büttner, C.X. Liu, G. Tkachov, E.G. Novik, C. Brüne, H. Buhmann, E.M. Hankiewicz, P. Recher, B. Trauzettel, S.C. Zhang, and L.W. Molenkamp, "Single valley Dirac fermions in zero-gap HgTe quantum wells," Nat. Phys. **7,** 418 (2011).
11. B. Lv, T. Qian, and H. Ding, "Angle-resolved photoemission spectroscopy and its application to topological materials," Nat. Rev. Phys. **1,** 609 (2019).
12. A. Inbar, J. Birkbeck, J. Xiao, T. Taniguchi, K. Watanabe, B. Yan, Y. Oreg, A. Stern, E. Berg, and S. Ilani, "The quantum twisting microscope," Nature **614,** 682 (2023).
13. M. F. Crommie, C.P. Lutz and D.M. Eigler, "Imaging standing waves in a two-dimensional electron gas," Nature **363,** 524–527 (1993).
14. C. Zoth, P. Olbrich, P. Vierling, K.-M. Dantscher, V.V. Bel'kov, M.A. Semina, M.M. Glazov, L.E. Golub, D.A. Kozlov, Z.D. Kvon, N.N. Mikhailov, S.A. Dvoretsky, and S.D. Ganichev, "Quantum oscillations of photocurrents in HgTe quantum wells with Dirac and parabolic dispersions," Phys. Rev. B **90**, 205415 (2014).

15. Z.D. Kvon, S.N. Danilov, D.A. Kozlov, C. Zoth, N.N. Mikhailov, S.A. Dvoretskii, and S.D. Ganichev, "Cyclotron resonance of Dirac ferions in HgTe quantum wells," JETP Letters **94,** 816 (2012).

16. V. Dziom, A. Shuvaev, N.N. Mikhailov, and A. Pimenov, "Terahertz properties of Dirac fermions in HgTe films with optical doping," 2D Materials **4,** 24005 (2017).

17. J. Ludwig, Y.B. Vasilyev, N.N. Mikhailov, J.M. Poumirol, Z. Jiang, O. Vafek, and D. Smirnov, "Cyclotron resonance of single-valley Dirac fermions in nearly gapless HgTe quantum wells," Phys. Rev. B **89,** 241406 (2014).

18. A.M. Shuvaev, V. Dziom, N.N. Mikhailov, Z.D. Kvon, Y. Shao, D.N. Basov, and A. Pimenov, "Band structure of a two-dimensional Dirac semimetal from cyclotron resonance," Phys. Rev. B **96,** 155434 (2017).

19. S. Dvoretsky, N. Mikhailov, Y. Sidorov, V. Shvets, S. Danilov, B. Wittman, and S. Ganichev, "Growth of HgTe Quantum Wells for IR to THz Detectors," Journal of Elec. Materials **39**, 918 (2010).

20. S. T. Pantelides ed., *"Deep Centers in Semiconductors,"* (Gordon and Breach, New York, 1986).

21. P.M. Mooney and T.N. Theis, "The DX center: a new picture of substitutional donors in compound semiconductors," Comments on Cond. Matter Phys. **16**, 167-190 (1992).

22. S.D. Ganichev, "Intense terahertz excitation of semiconductors" (Oxford Univ. Press, Oxford, 2009).

23. P. Olbrich, C. Zoth, P. Vierling, K.-M. Dantscher, G.V. Budkin, S.A. Tarasenko, V.V. Bel'kov, D.A. Kozlov, Z.D. Kvon, N.N. Mikhailov, S.A. Dvoretsky, and S.D. Ganichev, "Giant photocurrents in a Dirac fermion system at cyclotron resonance" Phys. Rev. B **87**, 235439 (2013).

24. A. Ferreira, J. Viana-Gomes, Y.V. Bludov, V. Pereira, N.M.R. Peres, and A.H. Castro Neto, "Faraday effect in graphene enclosed in an optical cavity and the equation of motion method for the study of magneto-optical transport in solids," Phys. Rev. B **84,** 235410 (2011).

25. J.C. König-Otto, Y. Wang, A. Belyanin, C. Berger, W.A. de Heer, M. Orlita, A. Pashkin, H. Schneider, M. Helm, and S. Winnerl, "Four-Wave Mixing in Landau-Quantized Graphene," Nano letters **17,** 2184 (2017).

26. D.B. But, M. Mittendorff, C. Consejo, F. Teppe, N.N. Mikhailov, S.A. Dvoretskii, C. Faugeras, S. Winnerl, M. Helm, W. Knap, M. Potemski, and M. Orlita, "Suppressed Auger scattering and tunable light emission of Landau-quantized massless Kane electrons," Nat. Photon. **13,** 783 (2019).

27. T. Maag, A. Bayer, S. Baierl, M. Hohenleutner, T. Korn, C. Schüller, D. Schuh, D. Bougeard, C. Lange, R. Huber, M. Mootz, J.E. Sipe, S.W. Koch, and M. Kira, "Coherent cyclotron motion beyond Kohn's theorem," Nat. Phys. **12,** 119 (2016).

28. M. Borsch, C.P. Schmid, L. Weigl, S. Schlauderer, N. Hofmann, C. Lange, J.T. Steiner, S.W. Koch, R. Huber, and M. Kira, "Super-resolution lightwave tomography of electronic bands in quantum materials," Science **370,** 1204 (2020).

29. J. Mornhinweg, M. Halbhuber, C. Ciuti, D. Bougeard, R. Huber, and C. Lange, "Tailored Subcycle Nonlinearities of Ultrastrong Light-Matter Coupling," Phys. Rev. Lett. **126,** 177404 (2021).

30. E.G. Novik, A. Pfeuffer-Jeschke, T. Jungwirth, V. Latussek, C.R. Becker, G. Landwehr, H. Buhmann, and L.W. Molenkamp, "Band structure of semimagnetic $Hg_{1-y}Mn_yTe$ quantum wells," Phys. Rev. B **72,** 35321 (2005).

31. K.-M. Dantscher, D.A. Kozlov, P. Olbrich, C. Zoth, P. Faltermeier, M. Lindner, G.V. Budkin, S.A. Tarasenko, V.V. Bel'kov, Z.D. Kvon, N.N. Mikhailov, S.A. Dvoretsky, D. Weiss, B. Jenichen, and S.D. Ganichev, "Cyclotron-resonance-assisted photocurrents in surface states of a three-dimensional topological insulator based on a strained high-mobility HgTe film," Phys. Rev. B **92,** 165314 (2015).

32. A. Uri, S. Grover, Y. Cao, J.A. Crosse, K. Bagani, D. Rodan-Legrain, Y. Myasoedov, K. Watanabe, T. Taniguchi, P. Moon, M. Koshino, P. Jarillo-Herrero, and E. Zeldov, "Mapping the twist-angle disorder and Landau levels in magic-angle graphene," Nature **581,** 47 (2020).

33. C.-C. Lu, H.-Y. Yuan, H.-Y. Zhang, W. Zhao, N.-E. Zhang, Y.-J. Zheng, S. Elshahat, and Y.-C. Liu, "On-chip topological nanophotonic devices," Chip **1,** 100025 (2022).

34. T.A. Uaman Svetikova, T.V.A.G. de Oliveira, A. Pashkin, A. Ponomaryov, C. Berger, L. Fürst, F.B. Bayer, E.G. Novik, H. Buhmann, L.W. Molenkamp, M. Helm, T. Kiessling, S. Winnerl, S. Kovalev, and G.V. Astakhov, "Giant THz Nonlinearity in Topological and Trivial HgTe-Based Heterostructures," ACS Photonics **10,** 3708 (2023).

35. S. Ito, M. Schüler, M. Meierhofer, S. Schlauderer, J. Freudenstein, J. Reimann, D. Afanasiev, K.A. Kokh, O.E. Tereshchenko, J. Güdde, M.A. Sentef, U. Höfer, and R. Huber, "Build-up and dephasing of Floquet–Bloch bands on subcycle timescales," Nature **616,** 696 (2023).

36. M. Borsch, M. Meierhofer, R. Huber, and M. Kira, "Lightwave electronics in condensed matter," Nat. Rev. Mater. **8,** 668 (2023).

37. D. Choi, M. Mogi, U. de Giovannini, D. Azoury, B. Lv, Y. Su, H. Hübener, A. Rubio, and N. Gedik, "Observation of Floquet–Bloch states in monolayer graphene," Nat. Phys. https://doi.org/10.1038/s41567-025-02888-8 (2025).

38. A. Blais, S.M. Girvin, and W.D. Oliver, "Quantum information processing and quantum optics with circuit quantum electrodynamics," Nat. Phys. **16,** 247 (2020).

# Supplementary Material

## All-optical band structure reconstruction
## and onset of Landau quantization of Dirac fermions

J. Riepl[1,+], M. Aichner[1,2,+], N. N. Mikhailov[3], S. A. Dvoretsky[3], G. Budkin[4], S. Ganichev[1], C. Lange[5],

J. Mornhinweg[1,5,6,*] and R. Huber[1]

[1]*Dept. of Physics, University of Regensburg, 93040 Regensburg, Germany*

[2]*LSI, CNRS, CEA/DRF/IRAMIS, Ecole Polytechnique, Institut Polytechnique de Paris,*

*F-91120 Palaiseau, France*

[3]*A.V. Rzhanov Institute of Semiconductor Physics, Novosibirsk 630090, Russia*

[4]*A. F. Ioffe Insitute, Saint Petersburg, Russia*

[5]*Dept. of Physics, TU Dortmund University, 44227 Dortmund, Germany*

[6]*Harvard John A. Paulson School of Engineering and Applied Sciences, 02138 Cambridge, MA, USA*

*Corresponding author, jmornhinweg@seas.harvard.edu*

[+]*These authors contributed equally*

# Contents

## Optical doping

For doping our HgTe QW, we used a white LED (spectrum: see Supplementary Fig. 1) placed between the plastic film and the cryostat window. A specially designed electronic circuit allows to switch-on the LED for a given exposure time $t_\text{e}$.

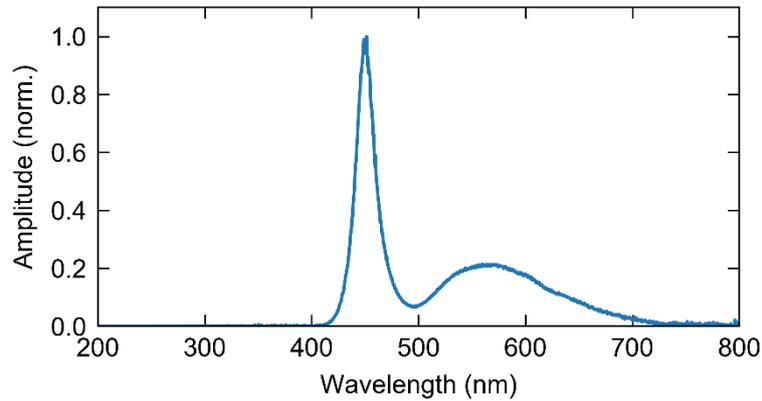

**Supplementary Figure 1. LED for optical doping.** Emission spectrum of the LED used to optically dope our HgTe QW.

This optical doping is based on the persistent photoconductivity effect. This effect is well-known in HgTe quantum well structures (see Refs. 1,2 for example). The most detailed study of optical doping was performed in Reference 2, which applied transport and cyclotron resonance measurements. It should be noted that the samples used in both of the cited works are similar to those used in our study. While the effect is clearly documented, the type of impurities resulting in optical ionization and persistent photoconductivity has not yet been explored. One might speculate that persistent photoconductivity is caused by the ionization of autolocalized impurities, such as DX and EL2 centers in III-VI compound semiconductors (e.g., AlGaAs heterostructures, see Refs. 3-5). In analogy to Refs. 4-5, Supplementary Figure 2 shows the adiabatic potential diagram representing an impurity with strong electron-phonon interaction and autolocalization.

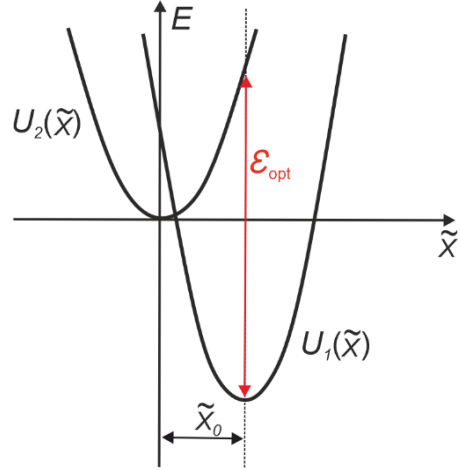

**Supplementary Figure 2.** Adiabatic potential diagram for an impurity with strong electron-phonon interaction and autolocalization. After Ref. 5.

The potential curves $U_1(\tilde{x})$ and $U_2(\tilde{x})$ correspond to the electron bound to the impurity and to the ionized impurity with zero kinetic energy of the electron, respectively. The equilibrium position of the bound state is shifted with respect to the ionized state due to the electron–phonon interaction. The energy separation between both potentials is determined by the electron binding energy $\mathcal{E}_b(\tilde{x})$ as a function of the configuration coordinate $\tilde{x}$. Taking into account the Franck–Condon principle, the bound state equilibrium energy yields the value of the threshold of optical ionization: $\mathcal{E}_{opt} = \mathcal{E}_b(\tilde{x}_0)$, where $\tilde{x}_0$ is the displacement of the bound state due to electron–phonon interaction. This diagram is used to describe, for instance, the DX- and EL2-centers, where this difference was experimentally revealed, see, e.g., Refs. 3,4,5. Such autolocalized states have a large potential barrier suppressing the return of free carriers to the localized state, thus giving rise to the phenomenon of persistent photoconductivity. Under these conditions, there is no radiative capture into the impurity state. While one might suppose that autolocalized impurities are present in the studied structure, to the best of our knowledge there are no publications describing experiments that study the impurities responsible for persistent photoconductivity.

# Sample structure

Our structure was grown on a (013)-oriented GaAs substrate using molecular beam epitaxy. First, a strain-compensating buffer layer of ZnTe and CdTe was deposited, followed by the QW system consisting of two $Hg_{0.35}Cd_{0.65}Te$ barrier layers (thickness, 39 nm) with the HgTe QW of the critical thickness of 6.5 nm sandwiched in between. The procedure of the growth buffer layer and the subsequent HgTe quantum well was described in Ref.6. In Supplementary Fig. 3 the growth layer schematic of the HgTe QW on a (013)-CdTe/ZnTe/GaAs complex substrate, which is grown in a single process by MBE, is shown.

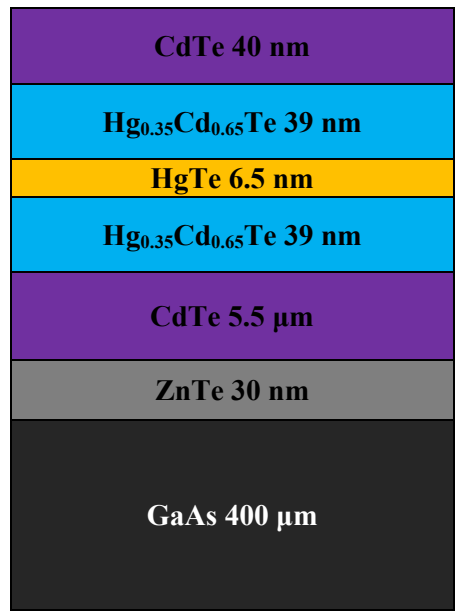

**Supplementary Figure 3.** Schematic of the layer distribution of the HgTe QW on a (013)-CdTe/ZnTe/GaAs substrate stack.

# Simulation of cyclotron resonances

In order to fit our measured cyclotron resonances and extract the underlaying band structure, we developed a distinct fitting procedure and a dielectric function describing the Landau-quantized HgTe-QWs.

## Hamiltonian of the system

As a starting point, we use the Hamiltonian for Landau-quantized HgTe-QWs developed in Ref. 7

$$\hat{H} = \mathcal{C}\mathbf{1} + \mathcal{M}\Gamma_5 - \frac{\mathcal{D}\mathbf{1} + \mathcal{B}\Gamma_5}{\hbar^2}(\hat{p} + e\mathbf{A}(\mathbf{r},t))^2 + \frac{\mathcal{A}\vec{\Gamma}}{\hbar}(\hat{p} + e\mathbf{A}(\mathbf{r},t)) + \frac{\mu_B B \Gamma_g^z}{2}. \quad (1)$$

The magnetic field is introduced within the principle of minimal coupling ($\hat{p} \to \hat{p} + e\mathbf{A}$). The $\Gamma$-matricies are given by

$$\Gamma_1 = \begin{pmatrix} \sigma_x & 0 \\ 0 & -\sigma_x \end{pmatrix}, \quad \Gamma_2 = \begin{pmatrix} -\sigma_y & 0 \\ 0 & -\sigma_y \end{pmatrix},$$
$$\Gamma_5 = \begin{pmatrix} \sigma_z & 0 \\ 0 & \sigma_z \end{pmatrix}, \quad \Gamma_g^z = \begin{pmatrix} \sigma_g & 0 \\ 0 & -\sigma_g \end{pmatrix}, \quad (2)$$

$$\vec{\Gamma} = (\Gamma_1, \Gamma_2), \quad (3)$$

where $\sigma_x$, $\sigma_y$ and $\sigma_z$ denote the Pauli-matrices and $\sigma_g = \text{diag}(g_e, g_h)$ contains the effective g-factors of the electron- (e) and heavy hole- (h) like states. For the rest of the fitting procedure we set $\mathcal{C} = 0$ and keep the Zeeman-Term in the Hamiltonian constant ($g_e = 60, g_h = 0$) at literature values[8] because our measurements are at magnetic field strengths lower or equal to 1.5 T.

The Hamiltonian without magnetic field

$$\hat{H} = \mathcal{C}\mathbf{1} + \mathcal{M}\Gamma_5 - \frac{\mathcal{D}\mathbf{1} + \mathcal{B}\Gamma_5}{\hbar^2}\hat{p}^2 + \frac{\mathcal{A}\vec{\Gamma}}{\hbar}\hat{p} \quad (4)$$

describes a two-band model which follows

$$E = \mathcal{C} - \mathcal{D}k^2 \pm \sqrt{\mathcal{M}^2 - 2\mathcal{M}\mathcal{B}k^2 + \mathcal{A}^2k^2 + \mathcal{B}^2k^4}$$
$$= \mathcal{C} - \mathcal{D}k^2 \pm \sqrt{(\mathcal{M} - \mathcal{B}k^2)^2 + \mathcal{A}^2k^2}. \qquad (5)$$

Therefore, our fitting procedure needs to fix the five free band structure parameters $(\mathcal{A}, \mathcal{B}, \mathcal{C}, \mathcal{D}, \mathcal{M})$.

**Pre-processing of experimental data**

The experimental data are smoothed using a Savitzky-Golay filter[9,10]. The Savitzky-Golay filter smooths data by applying a low-degree polynomial to values within a sliding window, using least squares fitting[9]. This technique is especially effective for reducing noise in data while preserving key features, making it valuable in preprocessing applications across fields such as spectroscopy[11,12,13,14]. The idea is to reduce the noise by replacing each data point with a local average of neighboring points. This is legitimate since nearby points in our measurement reflect nearly the same underlying value. Therefore, the averaging can effectively diminish noise without significantly distorting the true signal[9].

**Fit algorithm**

Our fit algorithm aims to test numerous data sets while conserving computational resources. Therefore, we employ a two-stage fitting procedure to identify the best-fit parameter set $\boldsymbol{p} = (\mathcal{A}^p, \mathcal{B}^p, \mathcal{C}^p, \mathcal{D}^p, \mathcal{M}^p, g_e^p, g_h^p)$. Initially, we preselect parameter sets, denoted as $M_0^0$, based solely on the requirement that the resonances must lie within the experimentally accessible measurement range. This preselection significantly reduces the number of candidate parameter sets but does not, by itself, allow for a precise determination of the best-fit parameters. Subsequently, for each parameter set $(\mathcal{A}, \mathcal{B}, \mathcal{C}, \mathcal{D}, \mathcal{M}, g_e, g_h, E_F)$ in $M_0^0$, we simulate the corresponding theoretical spectrum and compare it directly to the experimental data.

For the simulation, we use a transmission matrix formalism[15,16] to calculate the transmitted electric field through the quantum well. It requires the magnetic field-dependent dielectric function of the material $\epsilon_{ij}(\omega, B)$ as an input parameter in the three spatial directions $i, j = x, y, z$ (see below).

Due to the Dirac-like nature of the QW, the shape and structure of the cyclotron resonance with the characteristic discontinuities (see Fig. 3 main text) and square root $B$-field dependence (see equations (2) and (3) main text) contains significant information about the underlying bandstructure (see equation (5)). We fit the dielectric function, by varying the models' bandstructure parameters $(\mathcal{A}, \mathcal{B}, \mathcal{C}, \mathcal{D}, \mathcal{M})$, such that the simulation reproduces the experimental data as closely as possible. For different exposure times $t_e$, we only allow for different Fermi levels ($E_F$), while keeping the set of bandstructure parameters $(\mathcal{A}, \mathcal{B}, \mathcal{C}, \mathcal{D}, \mathcal{M})$ constant for all exposure times. Plugging these parameters into equation (5) allows us to draw conclusions about the band structure (in the limits of the model). We adjust the dielectric function's broadening factor $\Gamma$ to minimize deviation between simulated and recorded data. The temperature in the simulation is set to the experimentally recorded temperature. Accounting for the dielectric function's dependence on the Fermi energy introduces an additional parameter, such that our parameter set is $(\mathcal{A}, \mathcal{B}, \mathcal{C}, \mathcal{D}, \mathcal{M}, g_e, g_h, E_F)$.

Next, we simulate all parameter sets from the pre-selection $M_0^0$ and compare them with the experimental data, with the premise that there exists a parameter set $(\mathcal{A}, \mathcal{B}, \mathcal{C}, \mathcal{D}, \mathcal{M}, g_e, g_h)$, which describes all experiments well. Only $E_F$ varies depending on the doping time.

The best-fit parameter set is defined as the one whose simulated spectrum exhibits the minimal deviation $(\Delta_B, \Delta_\omega)$ from the experimental spectrum along both the magnetic field axis and the frequency axis. This approach is justified, as the spectral structure is fully determined by the transitions between Landau levels, which are, in turn, uniquely fixed by the chosen parameter set. Therefore, it suffices to compare the spectra at their defining edge points.

The refinement proceeds iteratively. We first establish maximum allowable deviations along both axes. We simulate spectra for all preselected parameter sets $M_0^0$ and compare them to the experimental spectrum recorded at the lowest exposure time (lowest experimental $E_F$). Parameter sets exceeding the predefined deviation thresholds are discarded. The remaining parameter sets are then compared with experimental spectra recorded at progressively higher exposure times. Therefore, an additional constraint is imposed: for a parameter set to be retained, the corresponding theoretical $E_F$ must increase consistently with the experimental $E_F$. Parameter sets failing this criterion are eliminated, and the surviving sets form a new subset, denoted $M_1^0 \subset M_0^0$. This procedure is repeated for all available experimental spectra, yielding a final reduced subset $M_n^0 \subset \cdots \subset M_0^0$, where $n$ corresponds to the number of experimental spectra considered. Once $M_n^0$ is obtained, the maximum allowable deviation $(\Delta_B, \Delta_\omega)$ is reduced. The procedure is then repeated, starting from $M_n^0$ (relabeled as $M_0^1$), until the final set contains the minimal number of parameter sets.

The remaining parameter set(s) after this iterative refinement define the best-fit parameter sets $P$, which in an ideal case contains, for different $E_F$, only one element, the best-fit parameter set $\boldsymbol{p} = (\mathcal{A}^p, \mathcal{B}^p, \mathcal{C}^p, \mathcal{D}^p, \mathcal{M}^p, g_e^p, g_h^p)$.

**Calculation of the dielectric function**

From our Hamiltonian (Eq. (1)), we derive the current operator as

$$\hat{j}_i = \hat{j}_{A_i^2} \mathbf{A}_i + \hat{j}_{P_i} + \hat{j}_{D_i} = \left(\frac{\mathcal{D}\mathbf{1} + \mathcal{B}\Gamma_5}{\hbar^2}\right)[2e(\hat{p} + e\mathbf{A})]_i + e\frac{\mathcal{A}}{\hbar}\vec{\Gamma}_i, \quad i \in 1,2. \tag{6}$$

where the first part represents the paramagnetic behavior, the second term the diamagnetic behavior and the last term accounts for the current operator of the Dirac-electrons[17]. Assuming spatial homogeneity, our focus shifts to the spatial two-dimensional average of the current operator $\hat{J} = \frac{1}{S}\int d^2r\, \hat{j}(r)$.

Using linear response theory, as detailed by Kubo, the first-order perturbation theory allows us to express the average current operator as:

$$\langle \hat{J}(t)\rangle = \langle \hat{J}(t)\rangle_{\hat{\rho}(0)} - \frac{i}{\hbar}\int_{t_0}^{t} dt'\, \langle [\hat{J}_I(t); \hat{H}_I(t')]\rangle_{\hat{\rho}(0)}, \tag{7}$$

where $\hat{H}_I(t)$ represents the light-matter-interaction (perturbation) Hamiltonian in the interaction picture. It is derived from the steady Hamiltonian as $\hat{H}(t) = (\hat{J}_P + \hat{J}_D)\mathbf{A}(\mathbf{r},t) = \hat{J}_{P,D}\mathbf{A}(\mathbf{r},t)$. Thus, the average current transforms into

$$\langle \hat{J}(t)\rangle = \langle \hat{J}_{A^2}\rangle_{\hat{\rho}(0)} - \frac{i}{\hbar}\int_{-\infty}^{\infty} dt'\, \Theta(t-t')\langle [\hat{J}_{P,D}(t)_I; \hat{J}_{P,D}(t')_I]\rangle_{\hat{\rho}(0)}\mathbf{A}(\mathbf{r},t). \tag{8}$$

Assuming equilibrium $\langle \hat{J}_{P,D}\rangle_{\hat{\rho}(0)} = 0$, no current flows.

To revert from the interaction picture to the Schrödinger picture, we utilize the unitary transformation $\hat{J}_S = \hat{U}\hat{J}_I\hat{U}^\dagger$ with $\hat{U} = \exp\left(-\frac{i}{\hbar}\hat{H}_0(t-t_0)\right)$. Considering $\hat{\rho}(0) = \frac{\exp(-\beta\hat{H}_0)}{Z}$ with $Z = \sum_n \exp(-\beta E_n)$, the expression for the $i$-th component of the average current becomes:

$$\langle \hat{J}_i(t)\rangle = \frac{1}{Z}\sum_n \exp(-\beta E_n)\langle \Psi_n|\hat{j}_{iA^2}|\Psi_n\rangle \mathbf{A}_j \delta_{ij}$$

$$-\frac{i}{\hbar}\int_{-\infty}^{\infty} dt'\, \theta(t-t')\exp(i(\omega_n - \omega_m)(t-t'))$$

$$\times \langle \Psi_n|\hat{j}_{i,P,D}|\Psi_m\rangle\langle \Psi_m|\hat{j}_{j,P,D}|\Psi_n\rangle(\exp(-\beta E_n) - \exp(-\beta E_m))\mathbf{A}_j(t'). \tag{9}$$

Choosing a monochromatic vector potential $\mathbf{A} = \mathbf{E}_0 \exp(i\omega t) + c.c$, and performing the time integral, we relate the current to conductivity in frequency space $\mathbf{J}(\omega) = \sigma(\omega)\mathbf{E}(\omega)$ arriving at the conductivity expression similar to that shown in Ref. 17, with an extra diamagnetic part:

$$\sigma_{ij}(\omega) = \frac{1}{i\omega S} \sum_n n_F(E_m)\langle\Psi_n|\hat{j}_{iA^2}|\Psi_n\rangle \delta_{ij}$$

$$-\frac{1}{i\omega S} \sum_{n\neq m} \left[\frac{(n_F(E_m) - n_F(E_n))}{(\omega + i\Gamma) + (\omega_m - \omega_n)} + (E_n \to -E_n \wedge E_m \to -E_m)\right]$$

$$\times \langle\Psi_n|\hat{j}_{i,P,D}|\Psi_m\rangle\langle\Psi_m|\hat{j}_{j,P,D}|\Psi_n\rangle. \tag{10}$$

Here we used the substitution of the Gibbs factors $Z^{-1}\exp(-\beta E_n)$ with the Fermi occupation number $n_F(E_n)$ shown in[17] and apply regularization to prevent infrared divergence[17]:

$$\sum_{n\neq m} \frac{1}{\omega}[\ldots] \to \sum_{n\neq m} \frac{1}{\omega_m - \omega_n}[\ldots] \tag{11}$$

This does not affect the resonance structure, as $\omega = \omega_m - \omega_n$ at resonance.

With Eq. (10) we evaluate the dielectric-function according to $\epsilon(\omega) = \epsilon_\infty + i\sigma(\omega)/(\epsilon_{2D} \cdot \omega)$, where we take into account that our sample is a quantum well with thickness $\delta_z$ and the 2D vacuum permittivity $\epsilon_{2D} = \delta_z \epsilon_0$. All background oscillation effects are accounted for by $\epsilon_\infty = 10$ (Ref. 18).

Next, we calculate the matrix-elements for the current-operator, adopting the eigenstates given for gauge $\mathbf{A}(\mathbf{r}) = -By\mathbf{e}_x$, as detailed in Ref. 7, denoting conduction/valence bands with $+$ and $-$.

$$\Psi^{\uparrow,\pm}_{n,k}(x,y) = \frac{\exp(ik_x x)}{\sqrt{L}\sqrt{l_B}}\begin{pmatrix} A_n^- \Phi_n(\epsilon) \\ A_n^+ \Phi_{n-1}(\epsilon) \\ 0 \\ 0 \end{pmatrix}, \Psi^{\uparrow,\pm}_{0,k}(x,y) = \frac{\exp(ik_x x)}{\sqrt{L}\sqrt{l_B}}\begin{pmatrix} \Phi_0(\epsilon) \\ 0 \\ 0 \\ 0 \end{pmatrix},$$

$$\Psi_{n,k}^{\downarrow,\pm}(x,y) = \frac{\exp(ik_x x)}{\sqrt{L}\sqrt{l_B}} \begin{pmatrix} 0 \\ 0 \\ B_n^+ \Phi_{n-1}(\epsilon) \\ B_n^- \Phi_n(\epsilon) \end{pmatrix}, \Psi_{0,k}^{\downarrow,\pm}(x,y) = \frac{\exp(ik_x x)}{\sqrt{L}\sqrt{l_B}} \begin{pmatrix} 0 \\ 0 \\ 0 \\ \Phi_0(\epsilon) \end{pmatrix}. \quad (12)$$

Where the coefficients are given as

$$A_n^\pm = \frac{\frac{\sqrt{2n}\mathcal{A}}{l_B} \mp \frac{\Delta_{\uparrow,n}}{2} \mp \left[\mathcal{M} - \frac{(2\mathcal{B}n + \mathcal{D})}{l_B^2} + \frac{(g_e - g_h)\mu_B B}{4}\right]}{\sqrt{\Delta_{\uparrow,n}(\Delta_{\uparrow,n} \mp 2\sqrt{2n}\mathcal{A}/l_B)}}, \quad (12)$$

$$B_n^\pm = \frac{-\left(\frac{\sqrt{2n}\mathcal{A}}{l_B}\right) \pm \frac{\Delta_{\uparrow,n}}{2} \mp \left[\mathcal{M} - \frac{(2\mathcal{B}n - \mathcal{D})}{l_B^2} + \frac{(g_e - g_h)\mu_B B}{4}\right]}{\sqrt{\Delta_{\uparrow,n}(\Delta_{\uparrow,n} \pm 2\sqrt{2n}\mathcal{A}/l_B)}} \quad (13)$$

with

$$\Delta_{\uparrow\downarrow,n} = 2\sqrt{\frac{2n\mathcal{A}^2}{l_B^2} + (\mathcal{M} - \frac{2\mathcal{B}n \pm \mathcal{D}}{l_B^2} \pm \frac{g_e - g_h}{4}\mu_B B)^2}. \quad (14)$$

Here the elements of the eigenfunctions are given by the parabolic cylinder functions $\Phi_n(\epsilon) = \frac{1}{\sqrt{n!\sqrt{\pi}}} D_n(\epsilon\sqrt{2}) = \frac{1}{\sqrt{n!\sqrt{\pi}}} \exp(-\epsilon^2/2) H_n(\epsilon)$ with $\epsilon = (\frac{y - k_y l_B^2}{l_B})$ and the following recurrence relations[19]:

$$\left(\frac{\epsilon'}{2} \pm \partial_{\epsilon'}\right) D_n(\epsilon') = \begin{cases} nD_{n-1}(\epsilon') \\ D_{n+1}(\epsilon') \end{cases}, \quad (15)$$

$$D_{n+1}(\epsilon') - \epsilon' D_n(\epsilon') + nD_{n-1}(\epsilon') = 0, \quad (16)$$

$$\int_{-\infty}^{\infty} d\epsilon' \exp(-\epsilon'^2) H_k(\epsilon') H_n(\epsilon') H_m(\epsilon') = \frac{2^s \pi^{\frac{1}{2}} \cdot k! \cdot n! \cdot m!}{(s-k)! \cdot (s-n)! \cdot (s-m)!}. \quad (17)$$

With $s = \frac{n+m+k}{2}$, for $n + m + k =$ even $\wedge$ $k, m, n \in \mathbb{N}$,

$$\int_{-\infty}^{\infty} d\epsilon' \exp(-\epsilon'^2) H_n(\epsilon') H_m(\epsilon') = 2^n \cdot n! \sqrt{\pi} \delta_{n,m}. \tag{18}$$

**Elements of the dielectric-function**

Next, we calculate the intra-band-transitions $n \mapsto n+1$, with the substitutions $A_n^\pm, B_n^\pm \cong X_n^\pm$ for the eigenstate coefficients.

The gauge choice for the vector field implies that our eigenstates are influenced by free wave propagation in the $x$-direction. Therefore, it is imperative to consider that each state is generally associated with a distinct $k_x$-state, denoted as $k$ or $k'$ in the eigenstates. This choice is somewhat arbitrary; we could also choose $\mathbf{A} = Bx\mathbf{e}_y$, which will lead to the expectation-values for $\hat{j}_{y_P}$ and $\hat{j}_{x_P}$ change their role. Thus, as long as we do not label any other axis except the axis of our magnetic-field, the dielectric-tensor should hold the symmetry-condition $\epsilon_{xx} = \epsilon_{yy}$. With these considerations it follows that the average of the current for $i,j \in \{x,y\}$ must hold the condition:

$$\langle \Psi_{n,k}^{\uparrow\downarrow} | \hat{j}_y | \Psi_{n+1,k'}^{\uparrow\downarrow} \rangle = i \langle \Psi_{n,k}^{\uparrow\downarrow} | \hat{j}_x | \Psi_{n+1,k'}^{\uparrow\downarrow} \rangle. \tag{19}$$

Here, the 90-degree rotation of $y$ introduces an additional factor of $i$. In further calculations, it is useful to recall that the momentum operator is Hermitian, satisfying the property $\langle \Psi | \hat{p}_i | \Psi \rangle = \langle \Psi | \hat{p}_i \Psi \rangle = \langle \hat{p}_i \Psi | \Psi \rangle$.

**Diamagnetic part**

$$\langle \Psi_{0,k}^{\uparrow\downarrow} | \hat{j}_{A^2} | \Psi_{0,k}^{\uparrow\downarrow} \rangle = \frac{2e^2 (\mathcal{D} \pm \mathcal{B})}{\hbar^2} \tag{20}$$

$$\langle \Psi_{n,k}^{\uparrow\downarrow} | \hat{j}_{A^2} | \Psi_{n,k}^{\uparrow\downarrow} \rangle = -\frac{2e^2 l_B \left[ (\mathcal{D} \pm \mathcal{B}) X_n^{-2} + (\mathcal{D} \mp \mathcal{B}) X_n^{+2} \right]}{\hbar^2} \tag{21}$$

**Paramagnetic part**

$$\langle \Psi_{0,k}^{\uparrow\downarrow} | \hat{j}_{y_P} | \Psi_{1,k'}^{\uparrow\downarrow} \rangle = i \frac{2e l_B}{\hbar} \frac{(\mathcal{D} \pm \mathcal{B}) X_1^-}{\sqrt{l_B}} \delta_{k,k'}, \tag{22}$$

$$\langle \Psi^{\uparrow\downarrow}_{n,k}|\hat{j}_{yP}|\Psi^{\uparrow\downarrow}_{n+1,k'}\rangle = i\frac{2el_B}{\hbar}\frac{[X_n^- X_{n+1}^-(\mathcal{D}\pm\mathcal{B}) + X_n^+ X_{n+1}^+(\mathcal{D}\mp\mathcal{B})]}{1}\delta_{k,k'} \qquad (23)$$

**Dirac part**

$$\langle \Psi^{\uparrow\downarrow}_{0,k}|\hat{j}_{yD}|\Psi^{\uparrow\downarrow}_{1,k'}\rangle = \pm i\frac{l_B}{\sqrt{l_B}}\frac{e\mathcal{A}}{\hbar}X_1^+\delta_{k,k'} \qquad (24)$$

$$\langle \Psi^{\uparrow\downarrow}_{n,k}|\hat{j}_{yD}|\Psi^{\uparrow\downarrow}_{n+1,k'}\rangle = \pm i\frac{el_B\mathcal{A}}{\hbar}X_n^- X_{n+1}^+\delta_{k,k'} \qquad (25)$$

Taking into account the conservation of momentum $k = k'$ given by $\delta_{k,k'}$, we obtain the conductivity:

$$\sigma_{ij}(\omega,B) = \frac{1}{2\pi l_B^2}\frac{1}{i\omega}\sum_n n_F(E_m)\langle\Psi_n|\hat{j}_{iA^2}|\Psi_n\rangle\delta_{ij}$$

$$-\frac{1}{2\pi l_B^2}\frac{1}{i\omega}\sum_{n\neq m}\left[\frac{(n_F(E_m) - n_F(E_n))}{(\omega + i\Gamma) + (\omega_m - \omega_n)} + (E_n \to -E_n \wedge E_m \to -E_m)\right]$$

$$\times \langle\Psi_n|\hat{j}_{i,P,D}|\Psi_m\rangle\langle\Psi_m|\hat{j}_{j,P,D}|\Psi_n\rangle. \qquad (26)$$

Where we used that the degeneration of each landau-level is given by the equation[17]

$\sum_k = (2\pi l_B^2)^{-1}$.

Finally, the complete dielectric-function tensor is given by:

$$\epsilon(\omega,B) = \begin{pmatrix} \epsilon_\infty & 0 & 0 \\ 0 & \epsilon_\infty & 0 \\ 0 & 0 & \epsilon_\infty \end{pmatrix} + \begin{pmatrix} \epsilon_{xx}(\omega,B) & \epsilon_{xy}(\omega,B) & 0 \\ \epsilon_{yx}(\omega,B) & \epsilon_{xx}(\omega,B) & 0 \\ 0 & 0 & \epsilon_{zz} \end{pmatrix}, \qquad (27)$$

with $\epsilon_{zz} = 10.8$ (Ref 20) and $\epsilon_\infty = 10$ (Ref. 18).

## k·p calculation

The electron spectrum is calculated using a **k·p** model, following well-established theoretical approaches[7,21]. The **k·p** calculations of the electron spectrum are carried out using the 6-band model, which consists of the conduction band ($\Gamma_6$) and the valence band ($\Gamma_8$). The system is described by an effective Hamiltonian $\widehat{H}$

$$\widehat{H} = \begin{pmatrix} \widehat{H}_c & \widehat{H}_{cv} \\ \widehat{H}_{cv}^\dagger & \widehat{H}_v + \widehat{H}_{BP} \end{pmatrix}. \qquad (28)$$

Block $\widehat{H}_c$ describes the conduction intraband contribution

$$\widehat{H}_c = I_{2\times 2}\left[E_c(z) + \frac{\hbar^2}{2m_0}\hat{k}(2F(z)+1)\hat{k} + \Xi_c \mathrm{Tr}(\varepsilon)\right] \qquad (29)$$

where $I_{2\times 2}$ is the identity matrix, $E_c(z)$ represents the conduction band profile, $k = (k_x, k_y, -i\partial/\partial z)$ is the wavevector operator, $F(z)$ accounts for remote band contributions, $\Xi_c$ is the $\Gamma_6$-band deformation potential, and $\varepsilon$ is the strain tensor. The coupling between the conduction and valence bands is described by the off-diagonal block $\widehat{H}_{cv}$, given by

$$\widehat{H}_{cv} = \begin{pmatrix} -\frac{1}{\sqrt{2}}Pk_+ & \sqrt{\frac{2}{3}}Pk_z & \frac{1}{\sqrt{6}}Pk_- & 0 \\ 0 & -\frac{1}{\sqrt{6}}Pk_+ & \sqrt{\frac{2}{3}}Pk_z & \frac{1}{\sqrt{2}}Pk_- \end{pmatrix} \qquad (30)$$

where $P$ is the Kane matrix element and $k_\pm = k_x \pm ik_y$. This coupling plays a crucial role in determining the band structure and optical properties of the system. The valence band is governed by the Luttinger Hamiltonian $\widehat{H}_v$, which describes the $\Gamma_8$ band structure and takes the form

$$\widehat{H}_v = E_v(z) + \frac{\hbar^2}{2m_0}\left[-\hat{k}\left(\gamma_1 + \frac{5}{2}\gamma_2\right)\hat{k} + 2\sum_\alpha J_\alpha \hat{k}_\alpha \gamma_2 J_\alpha \hat{k}_\alpha + 2\sum_{\alpha\neq\beta}\{J_\alpha, J_\beta\}_s \hat{k}_\alpha \gamma_2 \hat{k}_\beta\right], \qquad (31)$$

where $E_v(z) = E_c(z) - E_g(z)$ is valence band profile, $E_g(z)$ is band gap of respective layer, $\gamma_1$, $\gamma_2$ and $\gamma_3$ are the Luttinger parameters, $J$ is the vector of angular momentum 3/2 matrices, and $\{J_\alpha, J_\beta\}_s = (J_\alpha J_\beta + J_\beta J_\alpha)/2$ denotes symmetrized products. Strain effects in the valence band are incorporated via the Bir-Pikus Hamiltonian

$$H_{\text{BP}} = \left(a + \frac{5}{4}b\right)\text{Tr}(\varepsilon) - b\sum_\alpha J_{\alpha\alpha}^2 \varepsilon_{\alpha\alpha} - \frac{d}{\sqrt{3}}\sum_{\alpha \neq \beta} \{J_\alpha, J_\beta\}_s \varepsilon_{\alpha\beta} \tag{32}$$

where $a$, $b$ and $d$ are deformation potential constants for the $\Gamma_8$-band. The thick CdTe buffer is fully relaxed, while the HgTe quantum well and Cd$_{0.65}$Hg$_{0.35}$Te barriers inherit its in-plane lattice constant. The resulting lattice mismatch (~0.3 %) introduces strain in these layers, with the in-plane strain tensor components given by $\varepsilon_{xx} = \varepsilon_{yy} = a_{\text{CdTe}}/a_0 - 1$, where $a_{\text{CdTe}}$ is the CdTe lattice constant and $a_0$ is the unstrained lattice constant of the respective layer. For (001)-oriented structures, the out-of-plane strain $\varepsilon_{zz}$ is determined by minimizing elastic energy, yielding $\varepsilon_{zz} = -2\varepsilon_{xx} c_{12}/c_{11}$, with $c_{11}$ and $c_{12}$ being the elastic constants.

The calculations employ material parameters for HgTe and CdTe from Ref. 22, including bandgap energies $E_g$, Kane matrix elements $P$, Luttinger parameters $\gamma_1$, $\gamma_2$ and $\gamma_3$, deformation potentials $\Xi_c$, $a$, $b$, and $d$, and elastic constants $c_{11}$ and $c_{12}$, which are listed in Table 1. The valence band offset between HgTe and CdTe is set to 570 meV. The **k·p** model calculations based on this effective Hamiltonian enable computation of the electron spectrum in HgTe quantum wells, accounting for band mixing and strain effects. The calculated electron spectrum from this **k·p** model is presented in Fig. 4 (see main text).

|      | $E_g$ (eV) | $\frac{2m_0 P^2}{\hbar^2}$ (eV) | $F$ | $\gamma_1$ | $\gamma_2$ | $\gamma_3$ |
|------|------|------|------|------|------|------|
| HgTe | -0.303 | 18.8 | 0 | 4.1 | 0.5 | 1.3 |
| CdTe | 1.606 | 18.8 | -0.09 | 1.47 | -0.28 | 0.03 |

|      | $a$ (eV) | $b$ (eV) | $d$ (eV) | $\Xi_c$ (eV) | $a_0$ (Å) | $c_{11}$ (Mbar) | $c_{12}$ (Mbar) |
|------|------|------|------|------|------|------|------|
| HgTe | -0.13 | -1.5 | -8.0 | -3.82 | 0.646 | 0.597 | 0.415 |
| CdTe | 0.76 | -1.0 | -4.4 | -2.69 | 0.648 | 0.562 | 0.394 |

**Supplementary Table 1**: **Parameters of HgTe and CdTe used in k·p - model**

# References

1.  P. Olbrich, C. Zoth, P. Vierling, K.-M. Dantscher, G.V. Budkin, S.A. Tarasenko, V.V. Bel'kov, D.A. Kozlov, Z.D. Kvon, N.N. Mikhailov, S.A. Dvoretsky, and S.D. Ganichev, "Giant photocurrents in a Dirac fermion system at cyclotron resonance" Phys. Rev. B **87**, 235439 (2013).

2.  V. Dziom, A. Shuvaev, N.N. Mikhailov, and A. Pimenov, "Terahertz properties of Dirac fermions in HgTe films with optical doping," 2D Materials **4,** 24005 (2017).

3.  S. T. Pantelides ed., *"Deep Centers in Semiconductors,"* (Gordon and Breach, New York, 1986).

4.  P.M. Mooney and T.N. Theis, "The DX center: a new picture of substitutional donors in compound semiconductors," Comments on Cond. Matter Phys. **16**, 167-190 (1992).

5.  S.D. Ganichev, "Intense terahertz excitation of semiconductors" (Oxford Univ. Press, Oxford, 2009).

6.  S. Dvoretsky, N. Mikhailov, Y. Sidorov, V. Shvets, S. Danilov, B. Wittman, and S. Ganichev, "Growth of HgTe Quantum Wells for IR to THz Detectors," Journal of Elect. Materials **39**, 918 (2010).

7.  B. Scharf, A. Matos-Abiague, and J. Fabian, "Magnetic properties of HgTe quantum wells," *Phys. Rev. B* **86**, 075418 (2012).

8.  S. Gebert, C. Consejo, S.S. Krishtopenko, S. Ruffenach, M. Szola, J. Torres, C. Bray, B. Jouault, M. Orlita, X. Baudry, P. Ballet, S.V. Morozov, V.I. Gavrilenko, N.N. Mikhailov, S.A. Dvoretskii, and F. Teppe, "Terahertz cyclotron emission from two-dimensional Dirac fermions," *Nat. Photon.* **17**, 244 (2023).

9.  W.H. Press (ed), "Numerical recipes: The art of scientific computing," Cambridge University Press, Cambridge (2007).

10. P. Virtanen, R. Gommers, T.E. Oliphant, M. Haberland, T. Reddy, D. Cournapeau, E. Burovski, P. Peterson, W. Weckesser, J. Bright, S.J. van der Walt, M. Brett, J. Wilson, K.J. Millman, N. Mayorov, A.R.J. Nelson, E. Jones, R. Kern, E. Larson, C.J. Carey, İ. Polat, Y. Feng, E.W. Moore, J. VanderPlas, D. Laxalde, J. Perktold, R. Cimrman, I. Henriksen, E.A. Quintero, C.R. Harris, A.M. Archibald, A.H. Ribeiro, F. Pedregosa, and P. van Mulbregt, "SciPy 1.0: fundamental algorithms for scientific computing in Python," *Nat. Methods* **17,** 261 (2020).

11. F.S. Ruggeri, B. Mannini, R. Schmid, M. Vendruscolo, and T.P.J. Knowles, "Single molecule secondary structure determination of proteins through infrared absorption nanospectroscopy," *Nat. Commun.* **11,** 2945 (2020).

12. F.S. Ruggeri, J. Habchi, S. Chia, R.I. Horne, M. Vendruscolo, and T.P.J. Knowles, "Infrared nanospectroscopy reveals the molecular interaction fingerprint of an aggregation inhibitor with single Aβ42 oligomers.," *Nat. Commun.* **12,** 688 (2021).

13. J. Zheng, J. Nian, X. Ma, F. Zhang, and X. Qu, "Nonequal arm surface measurement of femtosecond optical frequency combs using the Savitzky-Golay filtering algorithm," *Applied optics* **61,** 9801 (2022).

14. J.-L.Z. Zaukuu, Z.S. Adams, N.A. Donkor-Boateng, E.T. Mensah, D. Bimpong, and L.A. Amponsah, "Non-invasive prediction of maca powder adulteration using a pocket-sized spectrophotometer and machine learning techniques," *Scientific reports* **14,** 10426 (2024).

15. I. Abdulhalim, "Analytic propagation matrix method for linear optics of arbitrary biaxial layered media," *Journal of Optics A: Pure and Applied Optics* **1,** 646 (1999).

16. I. Abdulhalim, "Analytic propagation matrix method for anisotropic magneto-optic layered media," *Journal of Optics A: Pure and Applied Optics* **2,** 557 (2000).

17. A. Ferreira, J. Viana-Gomes, Y.V. Bludov, V. Pereira, N.M.R. Peres, and A.H. Castro Neto, "Faraday effect in graphene enclosed in an optical cavity and the equation of motion method for the study of magneto-optical transport in solids," *Phys. Rev. B* **84,** 235410 (2011).

18. C.R. Becker, V. Latussek, A. Pfeuffer-Jeschke, G. Landwehr, and L.W. Molenkamp, "Band structure and its temperature dependence for type-III $HgTe/Hg_{1-x}Cd_xTe$ superlattices and their semimetal constituent," *Phys. Rev. B* **62,** 10353 (2000).

19. I.S. Gradštejn and J.M. Ryžik (eds), "Table of integrals, series and products," Elsevier Acad. Press, Amsterdam (2009).

20. T. Kernreiter, M. Governale, and U. Zülicke, "Quantum capacitance of an HgTe quantum well as an indicator of the topological phase," *Phys. Rev. B* **93** (2016).

21. E.G. Novik, A. Pfeuffer-Jeschke, T. Jungwirth, V. Latussek, C.R. Becker, G. Landwehr, H. Buhmann, and L.W. Molenkamp, "Band structure of semimagnetic $Hg_{1-y}Mn_yTe$ quantum wells," *Phys. Rev. B* **72**, 35321 (2005).

22. K.-M. Dantscher, D.A. Kozlov, P. Olbrich, C. Zoth, P. Faltermeier, M. Lindner, G.V. Budkin, S.A. Tarasenko, V.V. Bel'kov, Z.D. Kvon, N.N. Mikhailov, S.A. Dvoretsky, D. Weiss, B. Jenichen, and S.D. Ganichev, "Cyclotron-resonance-assisted photocurrents in surface states of a three-dimensional topological insulator based on a strained high-mobility HgTe film," *Phys. Rev. B* **92**, 165314 (2015).